\documentclass[structabstract]{aa}  
\usepackage{graphicx}
\usepackage{txfonts}
\begin{document}
   \title{Deep SDSS optical spectroscopy of distant halo stars}

   \subtitle{I. Atmospheric parameters and stellar metallicity distribution.}

   \author{C. Allende Prieto
          \inst{1,2}
          \and
          E. Fern\'andez-Alvar\inst{1,2}
      \and 
	  K. J. Schlesinger\inst{3}
      \and
          Y. S. Lee\inst{4}	  
      \and
          H. L. Morrison\inst{5}
      \and
          D. P. Schneider\inst{6}
      \and
          T. C. Beers\inst{7,8}
      \and
         D. Bizyaev\inst{9}
      \and 
         G. Ebelke\inst{9}
      \and
         E. Malanushenko\inst{9}
      \and
         V. Malanushenko\inst{9}
      \and
         D. Oravetz\inst{9}
      \and
          K. Pan\inst{9}
      \and
          A. Simmons\inst{9}
      \and J. Simmerer\inst{10}
     \and J. Sobeck\inst{11}
      \and A. C. Robin\inst{12}
          }

   \institute{Instituto de Astrof\'{\i}sica de Canarias,
              V\'{\i}a L\'actea, 38205 La Laguna, Tenerife, Spain\\              
         \and
             Universidad de La Laguna, Departamento de Astrof\'{\i}sica, 
             38206 La Laguna, Tenerife, Spain \\      
         \and
               Research School of Astronomy and Astrophysics, 
		The Australian National University, Weston, ACT 2611, Australia      \\
         \and
               Department of Astronomy, New Mexico State University, Las Cruces, NM 88003, USA \\
         \and
               Department of Astronomy, Case Western Reserve University,
		Cleveland, OH 44106, USA
         \and
             Department of Astronomy and Astrophysics, The Pennsylvania State University, 
             University Park, PA 16802, USA \\
         \and
 		National Optical Astronomy Observatory, Tucson, AZ, 85719, USA \\
	 \and
		JINA (Joint Institute for Nuclear Astrophysics), Michigan State University, East Lansing, MI 48824, USA 
         \and
             Apache Point Observatory, P.O. Box 59, Sunspot, NM 88349-0059, USA \\
         \and
             Lund Observatory, Box 43, SE-221 00 Lund, Sweden\\
	 \and
	    Department of Astronomy, University of Virginia, Charlottesville, VA 22903, USA
	 \and
	     Institut Utinam, CNRS UMR6213, Universit\'e de Franche-Comt\'e, 
	Observatoire de Besan\,con, Besan\,con, France \\
%                        \email{callende@iac.es}
%
             \thanks{Table1 is only available in electronic form at the CDS 
   via anonymous ftp to cdsarc.u-strasbg.fr (130.79.128.5) or via http://cdsweb.u-strasbg.fr/cgi-bin/qcat?J/A+A/}
             }

   \date{Received April 2014; accepted June 2014}

% \abstract{}{}{}{}{} 
% 5 {} token are mandatory
 
  \abstract
   {}
   {We analyze a sample of tens of thousands of spectra of
	halo turnoff stars, obtained with the optical spectrographs
	of the Sloan Digital Sky Survey (SDSS), 
	to characterize the  stellar halo population ``in situ''
	out to a distance of a few tens of kpc from the Sun. In this paper we describe
	the derivation of atmospheric parameters. We also derive the overall stellar
	metallicity distribution based on F-type stars observed as flux calibrators
	for the Baryonic Oscillations Spectroscopic Survey (BOSS).}  
   {Our analysis is based on an automated method that determines the set 
	of parameters of a model atmosphere that reproduces each 
	observed spectrum best. We used an optimization algorithm and evaluate
	model fluxes by means of interpolation in a precomputed grid.  
	In our analysis, we account for the spectrograph's varying resolution 
	as a function of fiber and wavelength. Our results for early
	SDSS (pre-BOSS upgrade) data compare well with those from the 
	SEGUE Stellar Parameter Pipeline (SSPP), except for stars with $\log$ g (cgs units) 
	lower than 2.5. 
}
   {An analysis of stars in the globular cluster M13 reveals a dependence of 
   	the inferred metallicity on surface gravity for stars with $\log g < 2.5$, confirming
   	the systematics identified in the comparison with the SSPP. We find that our metallicity 
    estimates  are significantly more precise than the SSPP results.
	We also find excellent agreement with several independent analyses.
	We show that the SDSS color criteria for 
	selecting F-type halo turnoff stars as flux calibrators 
	efficiently excludes stars with high metallicities, but does not 		significantly distort 
	the shape of the metallicity distribution at low
	metallicity. We obtain a halo metallicity distribution that is narrower
	and more asymmetric than in previous studies. The lowest gravity stars
	in our sample, at tens of kpc from the Sun, indicate a shift
	of the metallicity distribution to lower abundances, consistent with what is 
	expected from a dual halo system in the Milky Way.
	}
   {}

   \keywords{stars: abundances, fundamental parameters, population II -- 
		Galaxy: stellar content, halo
               }

   \maketitle
%
%________________________________________________________________

\section{Introduction}

The stellar halo population in the Milky Way was first noticed 
among solar neighborhood stars in the previous century
(Baade 1944;  Chamberlain \& Aller 1951), 
and as observational capabilities expanded, so did our knowledge 
about its spatial,  age, metallicity, and kinematical distribution.

Dedicated surveys of the halo, such as Beers, Preston, \& Shectman (1985, 1992), 
Norris et al. (1985), Ryan \& Norris (1991),  
Majewski (1992),  and Chiba \& Beers (2000),
have pioneered the sampling of halo stars at large distances.
More recently, cosmological surveys of galaxies and quasars, including
observations of stars for different purposes, have 
taken the lead in exploring the most distant parts of the halo.
The stars with the lowest known iron abundance come from candidates 
identified in the Hamburg-ESO survey (Wisotzki et al. 1996; 
Christlieb et al. 1999, 2002) 
or the Sloan Digital Sky Survey (SDSS; York et al. 2000, 
Bonifacio et al. 2012),  
and these studies constitute the best available 
database for studying the main properties of the stellar Galactic halo.

Most of the halo turnoff stars observed in the first phases of the SDSS
(Stoughton et al. 2002, Abazajian 2003, 2004) were targeted as flux calibrators, 
with the criterion $0.1< (g-r) <0.4$\footnote{
Here and throughout the rest of the manuscript,
it is understood that the magnitudes and colors are corrected
for the effects of absorption and reddening along the line of site,
based on the approach of Schlegel et al. (1998).
},
with slight variations over the years, and with additional significant numbers 
from SEGUE (Yanny et al. 2009), in particular their 
metal-poor MS turnoff and FG-type categories.

Among the four spectroscopic surveys included in the current phase  
of the Sloan Digital Sky Survey (SDSS-III; Eisenstein et al. 2011; 
Ahn et al. 2012),
two are dedicated to Galactic stars. It is, however, a third program
devoted to the study of the distributions of galaxy and quasar redshifts,
the Baryonic Oscillations Spectroscopic Survey (BOSS; Smee et al. 2012; 
Dawson et al. 2013), 
that provides the largest samples of distant Milky Way halo turnoff stars.

In this series of papers we use the spectrophotometric calibration
stars from BOSS and earlier  SDSS phases to examine the main
characteristics of the Galactic stellar
halo. Our sample constitutes the deepest
sample of halo turnoff stars and subgiants yet assembled. 
We describe the observations available
in \S \ref{observations} and the analysis in \S \ref{analysis},
followed by a description of the overall halo metallicity distribution, 
ending with a summary in Section \ref{conclusion}. Subsequent papers
will examine variations of halo-star chemistry with distance to the 
center of the Milky Way in detail.

\section{Observations}
\label{observations}

The spectra used in this work are included in SDSS Data Release
10 (DR10; Ahn et al. 2014) and come from the original 
SDSS spectrographs and their upgraded version, in operation since
2010 (Smee et al. 2012),
%(Eisenstein et al. 2011, Ahn et al. 2012) 
and the 2.5-m telescope 
(Gunn et al. 2006) at Apache  Point Observatory. They were 
obtained between the beginning of the survey (2000) and July 2012. 
Most quasars, galaxies, and stars observed spectroscopically in SDSS 
are chosen purely based on SDSS {\it ugriz} imaging (Fukugita et al. 1996;
 Stoughton et al. 2002; Eisenstein et al. 2011).

\subsection{Instrumentation}

The original SDSS spectrographs are two double spectrographs
covering between 380 and 920 nm with a FWHM resolving power
 $\lambda/\delta\lambda$ of about 2000, fed with 640 three-arcsecond 
optical fibers.

The BOSS spectrographs,  in operation since 2009, constitute an upgrade
of the original SDSS spectrographs. 
The new spectrographs have a higher multiplexing ability 
(1000 fibers in a single exposure instead of 640), and 
a wavelength-dependent resolving power in the range 
$1300 < \lambda/\delta\lambda< 3000$ between 360 and 1040 nm.
Their sensitivity is 
significantly enhanced from the original instruments, mainly by
adopting new detectors and volume-phase holographic gratings, as well as using
narrower (2 instead of 3 arcseconds in diameter) fibers that feed
light from the telescope. 
%This enhancement is immediately noticeable
%in the magnitude range of the flux standards in BOSS 
%($g<21$) and SDSS ($g<18.5$), despite a similar integration
%time and only slightly lower signal-to-noise ratios for BOSS.
%The spectrographs also offer an enhanced spectral coverage
%relative to their predecessors, from 360 to 1000 nm.
%It's unclear to me why fainter stars are used as spectrophotometric
%standards. If brighter ones are available, they should be preferred]

\subsection{Target selection}
\label{target}

We apply our analysis to the bulk of BOSS stellar spectra, which are 
defined as those with a measured redshift $|z|<0.01$, or a radial 
velocity under 3000 km s$^{-1}$, obtained as of late 2011. Some 
of these spectra, those obtained up to July 2011, became 
publicly available as of July 2012 in DR9 (Ahn et al. 2012). The
rest were made public in DR10 (Ahn et al. 2014).
We process all BOSS spectra exactly in the same manner as the observations
with the original SDSS spectrographs and they are described in \S \ref{analysis}, 
to derive the atmospheric
parameters ($T_{\rm eff}$, $\log g$, [Fe/H]). 

When available, we adopted the {\it Elodie} redshifts measured by the 
SDSS/SEGUE {\it spec1d} pipeline. These are obtained 
by template matching against a smooth version of spectra in the
Elodie library (Prugniel \& Soubiran 2001; Prugniel et al. 2007). 
When the {\it Elodie} redshifts
were not available ({\it elodie\_z} set to a value of zero), 
we embraced the standard {\it best}
redshift values adopted by the BOSS spectral pipeline (Bolton et al. 2012).

The sample of BOSS targets with nearly zero redshift we identify 
is mainly comprised of
\begin{enumerate}
\item F-type halo turnoff stars, those used for flux calibration;
\item Cooler (mostly K and M-type) stars, selected as part of
	ancillary science programs or by error;
\item White dwarfs, selected either by error or associated with
	ancillary science programs;
\item Hybrid spectra, showing a hot source in the blue (typically
	a white dwarf) and a cool source (low-mass star) dominating
	at redder wavelengths;
\item Galaxies or (mostly) quasars, with an incorrectly assigned redshift.
\end{enumerate}

Since SDSS/BOSS observations are mainly devoted to cosmology, 
most observations point to high
Galactic latitudes, avoiding the Galactic disk and favoring the halo
population. The exceptions are some of the SEGUE fields at low Galactic
latitude. With an absolute magnitude $M_g \sim 5$, F-type halo turnoff stars
observed by the SDSS/BOSS spectrographs 
are located at distances out to a few tens of kpc, 
and more evolved, horizontal-branch stars with similar temperatures 
are at distances in excess of 100 kpc.

When SDSS came into operation, the difficulty
of performing flux calibration for spectra having a wide wavelength 
coverage over of a large area of the sky was quickly recognized. 
Good standards were scarce 
and relatively bright, a situation that, although somewhat improved, 
persists today. The solution adopted for SDSS was to assign spectral 
types and corresponding model flux distributions to the stars, 
scale the model distributions to match the available five-band $ugriz$ SDSS 
photometry, and hope that the systematic errors are small and that random 
errors cancel out over the size of an SDSS spectroscopic plug-plate, 
 three degrees in diameter.

F-type halo turnoff stars, in the magnitude range spectroscopically 
explored by the original SDSS ($14<g<21$), are relatively abundant 
and fairly easy to model, because their continua are
shaped by a combination of H and H$^-$ bound free opacity, and 
they exhibit limited line absorption in the optical, mainly due to
iron lines. Taking BD $+17~4708$, with [Fe/H]$=-1.7$, 
as a prototype, the method was
developed and applied, with good results. The same protocol remains 
in use for BOSS, using somewhat fainter stars.

As in earlier phases of the SDSS, halo turnoff F-type stars are 
identified in BOSS observations by their colors. Typically
 16 of these stars are included per plugplate for flux calibration. 
These stars satisfy
\begin{eqnarray}
 15.0   & <r<      &  19     \nonumber \\
 u-g  & = 0.82  & \pm 0.08   \nonumber \\
 g-r  & = 0.30  & \pm 0.08  \nonumber \\
 r-i  & = 0.09  & \pm 0.08  \nonumber \\
 i-z  & = 0.02  & \pm 0.08,
\label{color_selection}
\end{eqnarray}
\noindent with slight variations used in the early phases
of the project. 
These stars are chosen based on the colors of the 
SDSS standard BD $+17$ 4708 (Oke 1990; Fukujita et al. 1996; 
Bohlin \& Gilliland 2004;  Ram\'{\i}rez et al. 2006); 
in addition, to satisfy the color and magnitude range in the
table, they are ranked in priority by their color resemblance   
to the standard star.

\section{Analysis}
\label{analysis}

Our analysis of the SDSS/BOSS spectra follows in its fundamentals
the methods described by Allende Prieto et al. (2006). However,
a number of improvements have been implemented. We describe the
basics and updates below.

\subsection{Analysis strategy and model spectra}

The BOSS spectrographs have a significant variation in the resolving
power as a function of wavelength. We account for such variations
by using the arrays of FWHM values that accompany the spectra 
and, through use of a grid of model spectra with higher resolution 
($\lambda/\delta\lambda=6400$), convolving the model spectra as we fit the observations
according to the FWHM for each wavelength and fiber.

The actual rest wavelengths of each BOSS spectrum differ from the
others in the same plugplate as a result of the stars' radial velocities. 
To avoid further interpolations in the observations, we resample the model
fluxes as we fit the data.
BOSS and earlier SDSS spectra are preprocessed with a custom 
IDL pipeline that rewrites the fluxes, wavelengths 
and line-spread functions for each spectrum in the format appropriate 
for FERRE, our FORTRAN90 fitting code (Allende Prieto et al. 2006, 2009). 
We performed tests  fitting either spectra with their
spectral energy distributions (suitably normalized by a constant to
make them independent of distance) or continuum normalized spectra, finding
that the latter method provides better results. Best-fitting parameters 
and spectra and estimated errors, were stored for further study.

   \begin{figure*}
   \centering
   \includegraphics[width=11cm,angle=90]{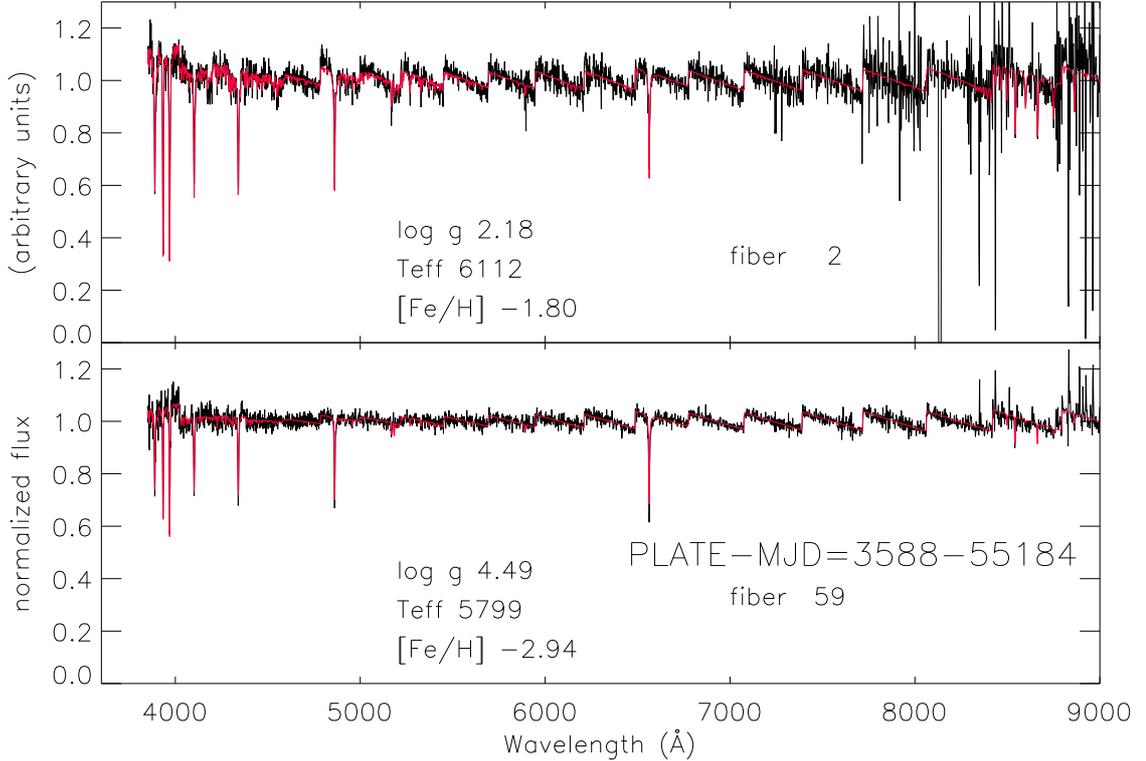}
      \caption{Sample BOSS spectra (black) for two representative low-metallicity 
	F-type stars, a dwarf and a giant.  The spectra are divided into 20 equal-velocity 
	bins and normalized by their mean value in each bin, as described in the text.
	The best-fitting parameters and models (red curves) are shown.}
         \label{spectra}
   \end{figure*}

Our method consists of finding the optimal set of model
atmosphere parameters that best matches the observed spectra.
We evaluate the fitness of the models with a simple $\chi^2$ 
criterion. For speed, model fluxes are precalculated on a regular
grid based on classical one-dimensional Kurucz model atmospheres,
and we interpolate in the three parameters under consideration 
($T_{\rm eff}$, $\log g$ and [M/H]) using
a quadratic Bezier scheme.

We make use of Kurucz model atmospheres (Castelli \& Kurucz 2004),
and the ASSET spectral synthesis code (Koesterke, Allende Prieto
\& Lambert 2008; Koesterke 2009). 
We adopt detailed continuum opacities from the compilation
by Allende Prieto et al. (2003) and subsequent updates (see, e.g., 
Allende Prieto 2008). Line data are
mainly from the calculations and literature compilations by Kurucz
(available from his website\footnote{kurucz.harvard.edu}), 
enhanced with damping constants from 
Barklem (2007 and references therein) when available.

We calculated a grid of model spectra covering $-5<$[Fe/H]$<+0.5$
and $0.5<\log g<4.5$ in steps of 0.5 dex, and $4750 <T_{\rm eff}<6500$
K, in steps of 250 K. The $\alpha$/Fe ratio adopted was solar
at [Fe/H]$=0$, and then linearly increasing toward lower [Fe/H],
reaching $+0.4$ for [Fe/H]$\le -1.5$, in agreement with the
typical values found for stars in the Milky Way\footnote{This
assumption will break down for some stars. See, e.g.,
Cohen et al. (2013).}. We adopted 
solar reference abundances as in Asplund et al. (2005), and
the overall metallicity of the model atmospheres was consistent
with those adopted in the spectral synthesis, with the exception
of the lowest metallicities, for which no model atmospheres were
readily available for our grid nodes. Model atmospheres with
[Fe/H]$=-2.5$ were adopted for [Fe/H]$=-3$, and those with
[Fe/H]$=-4$ were adopted for [Fe/H]$=-5.0$, $-4.5$, and $-4.0$.
The grid employed included 864 model spectra: 12 $\times 8 \times9$
([Fe/H]/$\log g$/$T_{\rm eff}$).

The parameters inferred from our analysis cannot be more accurate
than the models used to interpret the observations. Since 
we rely on large numbers of precalculated synthetic spectra, 
we cannot  consider aspects
such as departures from LTE or hydrodynamical (3D) effects. 
Nevertheless, we expect the impact of such a simplification 
on the derived  parameters will be quite uniform, given the limited range in
temperature and metallicity of the stars we consider.
As mentioned earlier, metal-poor F-type stars are one of the
simplest spectral types to work with.

Despite the BOSS spectrographs offering an increased spectral range
over the earlier SDSS spectrographs, we decided to adopt a
wavelength region common for the two data sets. This decision
was motivated by the need to evaluate the quality of our derived
parameters and the availability of parameters for large numbers
of SDSS stars from the  SEGUE (Yanny et al. 2009) 
Stellar Parameter Pipeline 
(hereafter SSPP; Lee et al. 2008a,b; Allende Prieto 2008) 
and a desire to combine SDSS-I, SEGUE, and BOSS observations.

Figure \ref{spectra} shows two examples of observed BOSS spectra, and 
our best-fitting models.  At very low metallicity, the 
information on the stellar properties
comes mainly from hydrogen (Balmer and Paschen) lines, Ca II lines 
(H and K in the blue, as well as the IR triplet). 
Our continuum correction scheme splits the spectral range
into 20 equal-velocity  bins and divides each by its mean value. This approach
removes large-scale systematic errors in flux, preserves some of the local
information on the continuum shape, and since it is a linear transformation,
it is robust (symmetric) against noise. This is an unusual continuum 
normalization procedure, but extensive tests with simulations led us 
to conclude that it is very robust and performs better than other, 
more commonly used methods for this kind of spectra. Exactly the same 
procedure is applied to both models and observations.

\subsection{Validation of parameters against SDSS/SEGUE}
\label{sspp}

Stellar spectra taken with the SDSS spectrographs before
the BOSS upgrade benefit from the added value of the 
SSPP. This software suite derives atmospheric parameters
([Fe/H]/$\log g$/$T_{\rm eff}$ and [$\alpha$/Fe]) for most
normal stars, and it has  been continuously upgraded 
(Lee et al. 2011, Smolinski et al. 2011). The SSPP offers an 
excellent comparison to our analysis, since
our techniques can be equally applied to BOSS or earlier SDSS
spectra.

The SSPP includes a wide variety of techniques to derive
each atmospheric parameter. The stellar effective temperature
is the one that admits more variants,  from Balmer-line equivalent
widths, to colors, line ratios, or spectral fitting -- about
ten different methods are considered by the SSPP. Metallicity is also
measured by a number of techniques, while gravity is the most
difficult parameter to extract, and only a handful of methods 
are available as part of the SSPP. Only a few of the techniques derive 
all parameters simultaneously and in a consistent manner.

An earlier version of our fitting code is 
included in the SSPP. However, it is only one of many, and
the version in the SSPP uses earlier grids of model fluxes,
limited to a narrow spectral range (440-550 nm). The main 
differences between our analysis and the SSPP are:
\begin{itemize}
\item We use only a single method, avoiding 
the complex problem of combining multiple techniques with
different performance levels over the parameters space;

\item We use a large fraction of the spectral range available 
from the SDSS spectrographs;

\item We account for the varying resolution of the spectra
with wavelength and fiber.
\end{itemize}

\begin{figure*}
   \centering
   \includegraphics[width=13cm]{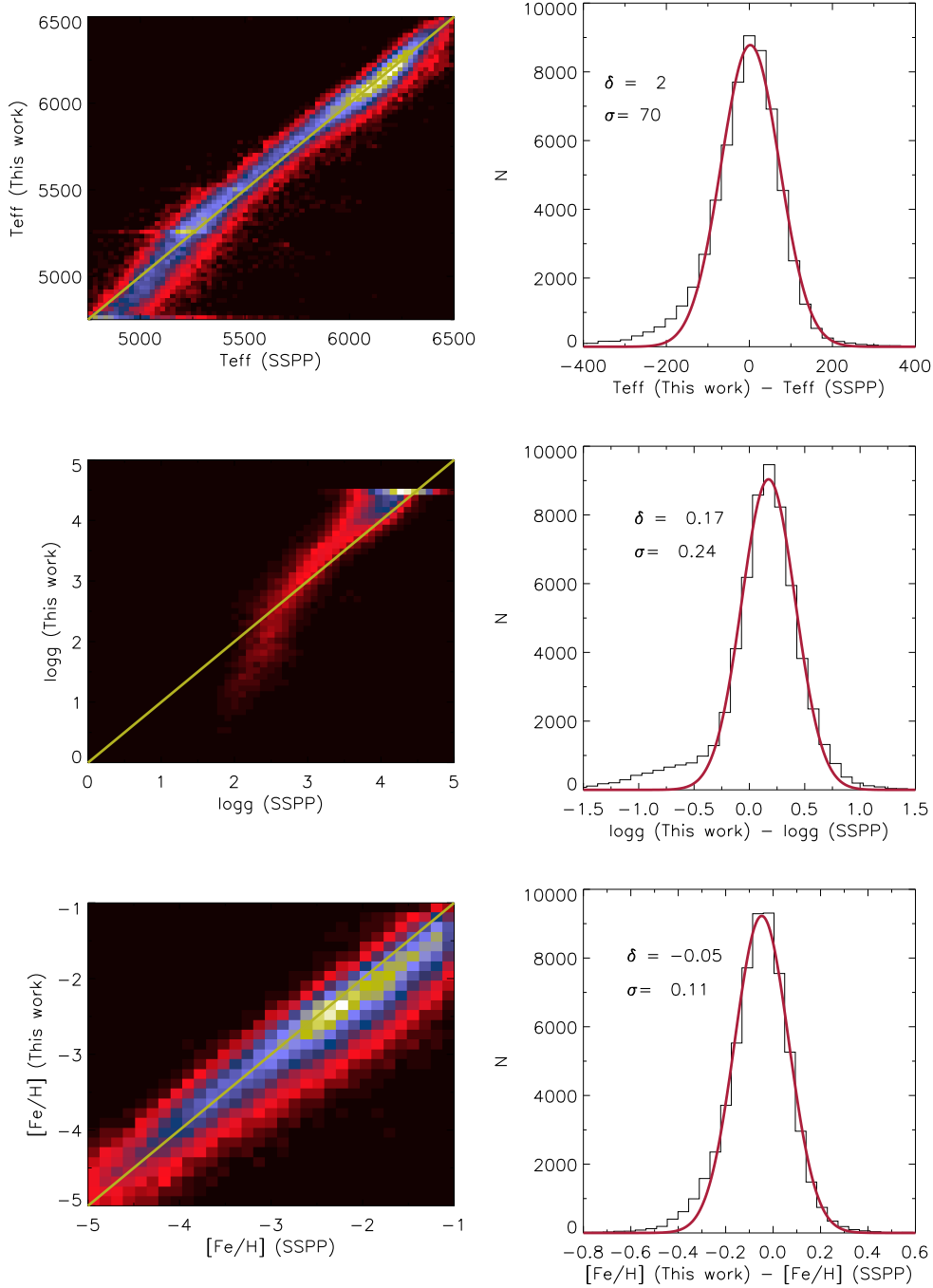}
      \caption{Comparison between the parameters derived by our method and those from 
      the Segue Stellar Parameter Pipeline (SSPP) for 61,040 F- and G-type stars in DR8. 
      The lefthand panels show the density distribution of stars in a logarithmic scale, 
      with the one-to-one relationship shown in yellow. 
      The righthand panels show the distribution of differences between this work
      and the SEGUE SSPP for each parameter, with Gaussian fittings and their 
      central values in red.}
         \label{dr8}
   \end{figure*}

We have processed all spectra in DR8 (Aihara et al. 2011) 
showing low redshifts ($cz<3000$ km s$^{-1}$)
with our method. This selection criterion identified 545 343 spectra.
We further trimmed this control sample by selecting objects that
we estimated to have an average signal-to-noise ratio per 
pixel\footnote{One pixel spans 69 km s$^{-1}$ in velocity.} 
between 400 and 800 nm higher than 30. They  
were fit by our models with a reduced $\chi^2<10$, that both our method and
the SSPP find to be in the range $4750<T_{\rm eff}<6500$ K, and with
metallicities [Fe/H]$<-1$, in order to eliminate the bulk of the 
Milky Way disk system. These criteria returned a sample of 61 040 stars, 
for which we have compared our atmospheric parameters and those from 
the SSPP. The results are illustrated in Fig. \ref{dr8}.

The agreement between the SSPP effective temperatures and metallicities in this
range of atmospheric parameters is excellent. Gaussian fittings to the 
distribution of the differences between this work and the SSPP are
$2 \pm 70$ K, $0.17\pm 0.24$ dex, and $-0.05 \pm 0.11$ dex for
$T_{\rm eff}$, $\log g$, and [Fe/H], respectively (see lefthand panels in Fig. 2).
%If we assume that each 
%analysis shares similar contributions
%to the random uncertainties in effective temperature and metallicity, 
%those must be, on average, 50 K and 0.08 dex, respectively, with
%no significant zero-point offsets. 

\begin{figure*}
   \centering
   		\includegraphics[width=7.2cm]{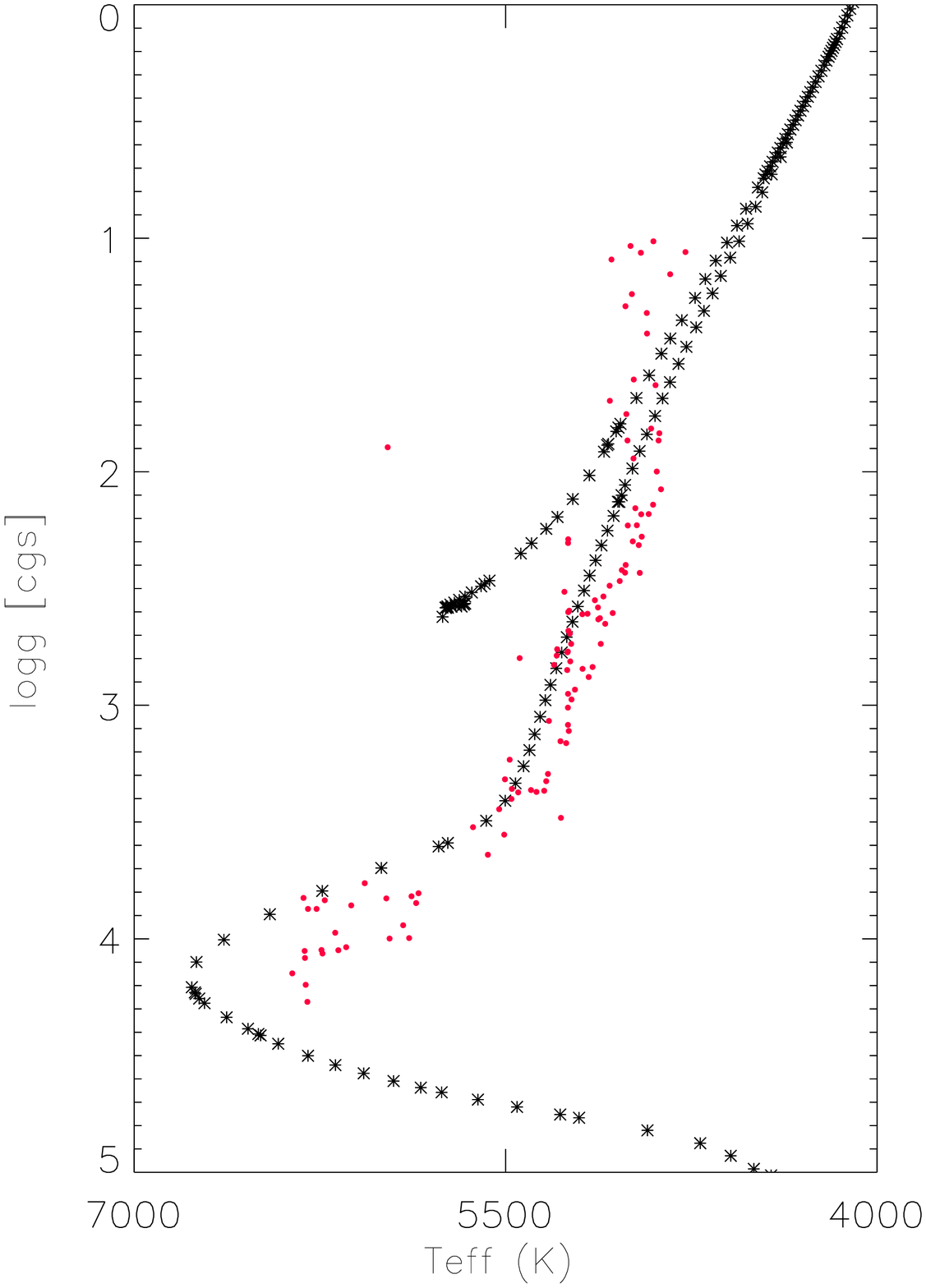}
   		\includegraphics[width=7.2cm]{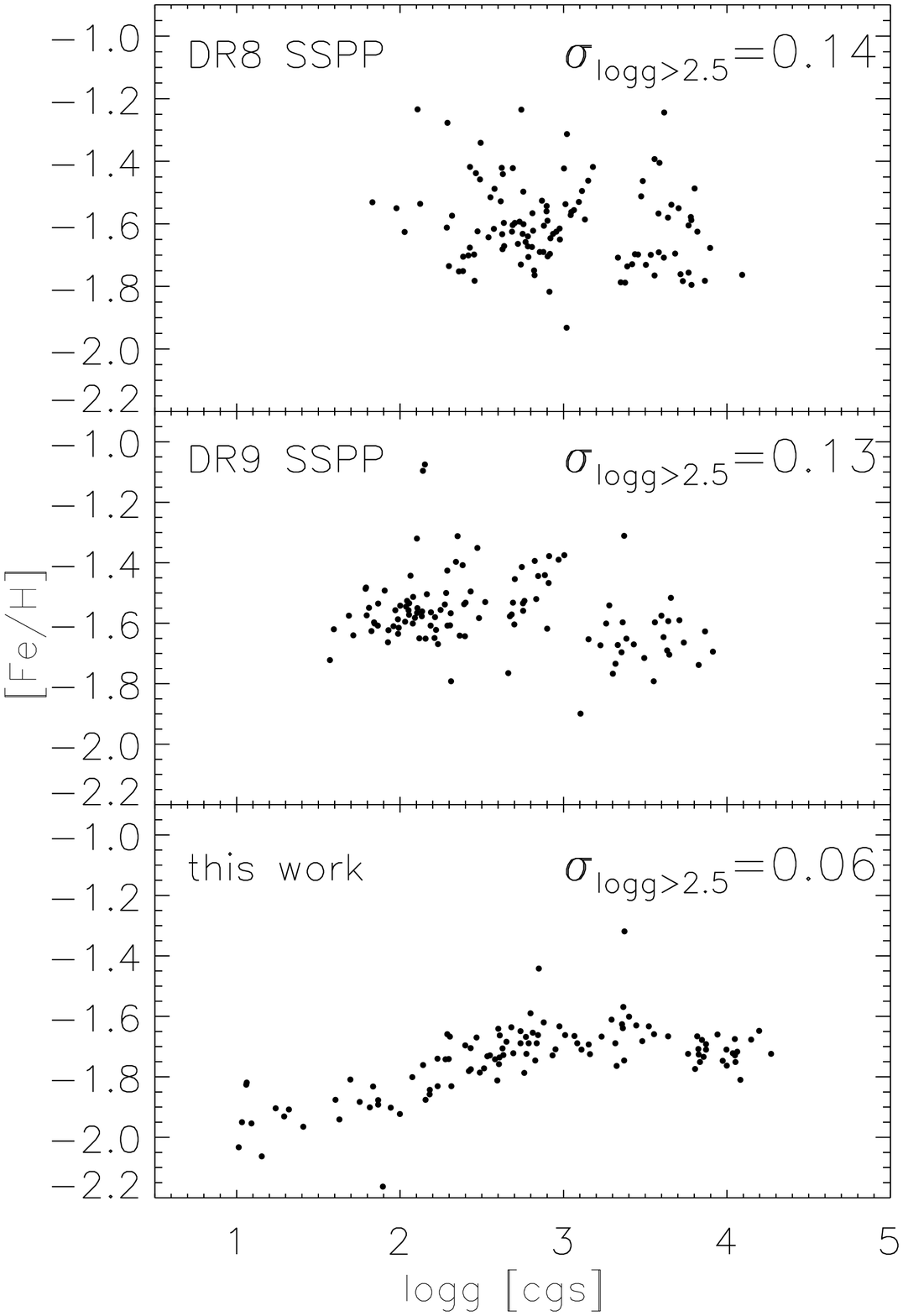} 		
   \caption{Analysis of the stars belonging to the globular cluster M13. 
   The left panel compares the stars (red dots) with an isochrone (black asterisks) 
   at 10.0 Gy and Z=0.0004 from Girardi et al. (2002).
   %\footnotetext{http://pleiadi.pd.astro.it/$\sim$lgirardi/isoc$\_$photsys.html}. 
   In the right panel [Fe/H] vs $\log g$ values from the SSPP in DR8 and DR9 are 
   plotted at the top; the bottom panel is the same plot for our results, which 
   shows evidence of a systematic decrease in the metallicity with the surface 
   gravity at $\log g < 2.5$ but a significant improvement in precision 
   over the SSPP results.}
   \label{m13}
   \end{figure*}

The agreement is not as good for surface
gravity, with our values slightly higher for dwarfs 
by about 0.15 dex, and 
significantly lower for giants ($\log g<2.5$), with discrepancies reaching
up to 1 dex at $\log g \simeq 1$. The surface gravities from the SSPP have been
throughly tested against globular and open clusters (Lee et al. 2008b, 
Smolinski et al. 2011) for 
gravities as low as $\log g \simeq 1.7$, and
therefore we suspect our values for low-gravity stars.
Examining the stars of this sample which belong 
to the globular cluster M13 clearly reveals a correlation for the metallicity 
with gravity at $\log g < 2.5$, as illustrated in Fig. \ref{m13}. 
Thus, the metallicity estimations for stars at this range are 
systematically underestimated owing to the errors in surface 
gravity. That leads us to be cautious with those parameters 
derived at $\log g < 2.5$, which will be excluded in our subsequent
analyses. This effect could at least be partly related to the fact
that low surface gravity red giants have temperatures that are close to our grid limit.
%In Section \ref{results} we will 
%derive the metallicity distribution function at different cuts 
%in $\log g$ and magnitude $g$ in order to constrain distances to the stars; 
%we will take into account only those at the confident range in surface gravity. 
On the other hand, the dispersion observed in our inferred metallicities 
for M13 stars with $\log g > 2.5$ is $\sigma=0.06$ dex, significantly 
lower than that corresponding to both DR8 and DR9 SSPP metallicity 
values for the same stars, $\sigma=0.14$ and $\sigma=0.13$, respectively. 

The model grid is limited to $\log g \le 4.5$
due to the inclusion of very low metallicities, for which no model atmospheres
for stars of higher gravity are available in the Castelli \& Kurucz (2004)
grids we employ. 
%In this paper we are concerned with distant halo
%stars, which have gravities $\log g < 4$, and therefore this limitation is not an issue.
We have built grids limited to [Fe/H]$\ge -2.5$ that 
reach $\log g = 5$, but
are otherwise identical, for other 
analyses presented in the following papers in this series.

As explained above, although BOSS spectra have a broader 
spectral range than the one provided
by the SDSS spectra in DR8, we limit our analysis to the range 380-900 nm,
common to all data. Thus, we expect that our confidence on
the metallicities we derive for metal-poor stars in DR8 can be 
extended to stars observed by BOSS with similar properties and 
signal-to-noise ratios.

DR9 involved some changes in the SSPP, mainly that the effective temperatures
were calibrated to match the infrared flux method scale (see, e.g., Casagrande et al. 2010). 
Repeating the previous comparison 
but adopting the SSPP DR9 parameters we find a very similar agreement,
although systematic offsets are more prominent than in DR8.
The Gaussian fittings to the 
distribution of the differences between this work and the SSPP are
$-54 \pm 66$ K, $0.23\pm 0.23$ dex, and $-0.11 \pm 0.11$ dex for
$T_{\rm eff}$, $\log g$, and [Fe/H], respectively.

\subsection{Validation against individual-line analysis of BOSS spectra}

A second method investigating the accuracy of the metallicities obtained in the 
analysis of BOSS spectra was implemented. This analysis focused on 
the  Ca II (H and K) resonance  lines (centered at about 397 and 393 nm), 
which saturate at high metallicity, 
but are a useful metallicity proxy at [Fe/H]$<-1$. Similar to our 
primary algorithm, we  determine [Fe/H] by comparing the $\chi^{2}$ between 
 observations and model spectra, but this time using only 
wavelengths in the vicinity of the Ca II lines, confining the comparison 
to the $T_{\rm eff}$ and $\log g$ values previously obtained, and
varying  [Fe/H]. A preliminary $\chi^{2}$ 
minimum is found by fitting three points with a parabola. 
We then evaluate the $\chi^2$ at five additional values of [Fe/H]
 around the minimum ($\pm 0.2$ dex and $\pm 0.4$ dex), then
fitting a new parabola to the constrained region around the minimum.

This method relies strongly on the fact that metal-poor stars
have enhanced Ca/Fe ratios relative to solar proportions 
and will  fail when
this is not true, or when the enhancement is far from the assumed
value ($+0.4$ at [Fe/H]$\le -1$), but this caveat also applies  
to our main method.  We derive an estimate of the uncertainty
in the metallicity associated with the noise in the spectrum, although
this ignores the error covariances with $T_{\rm eff}$ and $\log g$, since
these are held constant. We tested approximately a thousand stars
with metallicities between $-3<$[Fe/H]$<-1$ and found 
 excellent agreement with our main algorithm, with a mean metallicity
about $-0.09$ dex lower, and an rms deviation of 0.11 dex.
An attempt to consider both the Ca II resonance lines and the
Ca II IR triplet at about 860 nm provided slightly worse results.

\subsection{Additional validations}

A number of recent studies from the literature have investigated
metal-poor stars from SDSS. We can use these to obtain an
additional constraint on the accuracy of our atmospheric parameters.

As part of a larger sample Allende Prieto et al. (2008) analyzed 
a homogeneous set of high-resolution spectra  
from HRS on the 10-m Hobby-Eberly Telescope (Tull et al. 1998)
for 81 stars that had been previously observed with the SDSS spectrographs. 
Most of these stars have higher metallicities than those we 
are concerned with, but nevertheless
we find good agreement between our derived parameters 
for the 74 stars in the sample that 
are cooler than 6500 K (the warmest temperature we include in our sample) 
and the parameters published by Allende Prieto et al. (2008).
Our metallicities are on average 0.1 dex lower, our temperatures
are 130 K warmer, and our gravities are 0.25 dex higher than those
from the high-resolution analysis, with rms dispersions between 
the results of the two analyses (measured by fitting 
Gaussians to the residuals) 
of 0.18 dex, 163 K, and 0.23 dex in [Fe/H], $T_{\rm eff}$, and $\log g$, 
respectively.

Bonifacio et al. (2012)  present high-resolution VLT/UVES 
(Dekker et al. 2000) observations of 16 metal-poor stars with 
[Fe/H]$<-3$ identified from SDSS DR6 (Adelman-McCarthy et al. 2008). 
We have analyzed the SDSS spectra of these stars and compared our 
derived parameters with those from the analysis of high-resolution
spectra published by Bonifacio et al. Ignoring one star 
(SDSS J002113-005005) that is warmer than the models in our grid 
($T_{\rm eff}\sim$ 6550 K), the agreement is excellent. Our 
effective temperatures are on average  63 K higher, with an
rms scatter between the two sets of results of 114 K. Similarly,
our gravities are, on average, 0.13 dex higher with an rms 
scatter of 0.46 dex, and our metallicities are 0.3 dex higher
with an rms scatter of just 0.2 dex.

The same team of authors has recently announced the discovery of a dwarf
star with a metallicity around [Fe/H]$\sim -5$, SDSS J102915+172927,
originally identified as a candidate for ultra-low metallicity 
from the pool of SDSS spectra and later analyzed with higher
resolution data from the X-shooter and UVES spectrographs, both on the VLT. 
Our analysis of the SDSS spectrum of this relatively {\it bright} star for our sample 
($g=17.3$) yields estimates of 
$T_{\rm eff}= 5888$ K, $\log g=4.47 $, and [Fe/H]$=-4.4$, 
which compare well with those from the high-resolution 
analysis by Caffau et al. (2012), namely $T_{\rm eff}= 5811 \pm 150$ K, 
$\log g= 4.0 \pm$ 0.5, and [Fe/H]$= -4.89 \pm 0.10$.

An investigation by Yong et al. (2013), using several
high-resolution spectrographs, includes results on four stars 
originally observed with the SDSS spectrographs. We compared
our measurements for three of the stars, for which the team has derived
the three atmospheric parameters; our temperatures
are an average of 9 K cooler with an rms scatter of 88 K,
our surface gravities are on average 0.02 dex higher
with an rms scatter of 0.57 
dex, and our metallicities are an average of 0.34 dex higher with an rms
of 0.25 dex. For the fourth SDSS star (SDSS J014036+234458), 
Yong et al. derived
the effective temperature and metallicity assuming
that the star was either a dwarf or a giant, finding
a $T_{\rm eff}=5703$ K and [Fe/H]$=-4.0$ or $-4.1$, respectively.
Our analysis of the SDSS spectrum returned $T_{\rm eff}=5958$ K,
[Fe/H]$=-3.4$, and $\log g=3.8$.

Aoki et al. (2013) have published atmospheric parameters and compositions
for 137 stars, most of them on the turnoff, 
selected for having very low metallicities from SDSS/SEGUE. 
From 111 stars in common, we find that our $T_{\rm eff}$, 
$\log g$, and [Fe/H] values differ an average of $-99$ K ($\sigma=212$ K), 
$-0.01$ dex ($\sigma= 0.62$ dex),  and $+0.13$ dex ($\sigma=0.34$ dex), 
respectively, from theirs. This comparison adds very little to
the discussion about surface temperatures and gravities 
in \S \ref{sspp}, since these authors embrace the 
effective  temperatures provided by the SSPP and assume a constant 
$\log g$ of 4.0 with an expected uncertainty of about 0.5 dex 
for their turnoff stars. The large scatter in $T_{\rm eff}$ is driven by
a number of outliers at low temperature ($T_{\rm }< 5000$ K); 
a robust scatter estimate from a Gaussian model, as performed in \S \ref{sspp}, 
would instead show a mean offset of $-27$ K and a $\sigma$ of 93 K, in fair
agreement with the spread shown in the top panel in Fig. \ref{dr8}. 
Of relevance is the modest offset and scatter in metallicity, 
which Aoki et al. derive from Fe I lines in high-resolution
Subaru/HDS spectra (Noguchi et al. 2002). 
This scatter is about three times larger than the 
overall figure of $\approx$ 0.1 dex in Fig. \ref{dr8}, at least partly because 
most of the Aoki sample spans the range $-3.5<$[Fe/H]$<-2.5$; 
i.e.,  they have about one full dex lower metallicity than the
median of the sample considered in Fig. \ref{dr8}.

Overall, all these comparisons provide high confidence 
in our analysis of SDSS spectra of F and early G-type metal-poor 
stars and, in particular, on the inferred metallicities.

\section{The halo metallicity distribution function}
\label{results}

For our study of halo 
stars we selected those targeted in BOSS as spectrophotometric standards.
These are labeled as {\tt SPECTROPHOTO\_STD} by the pipeline, and they
represent a homogeneously selected sample of metal-poor stars. 

We limited the surface gravity to $2.5<\log g<4$, and $g>17$. All else being
equal, stars with lower gravities are more extended and luminous, and they probe
larger distances. Using stellar evolution models, we find that the above
magnitude and gravity limits
place our sample at distances beyond 5 kpc, restricting it mainly
to a halo population. We further limited our sample to stars fit
with a $\chi^2< 3$ and  a $T_{\rm eff}>5600$ K. For our 
selection of BOSS plugplates, this sample comprises about 5100 stars. 
If we instead consider a more relaxed upper limit in surface gravity,
$2.5<\log g<4.4$, our sample includes about 16 000 stars. 
We checked that imposing an additional selection on the average signal-to-noise
ratio per pixel to be higher than 30 did not alter our results significantly.
Repeated observations are about $12$\%, and therefore their effect 
on our analysis is negligible, but they provide a consistency check.
We find a mean rms for matched pairs of 73 K, 0.29 dex, and 0.16 dex 
in $T_{\rm eff}$, $\log g$, and [Fe/H], respectively.

   \begin{figure}
   \centering
   \includegraphics[width=8cm]{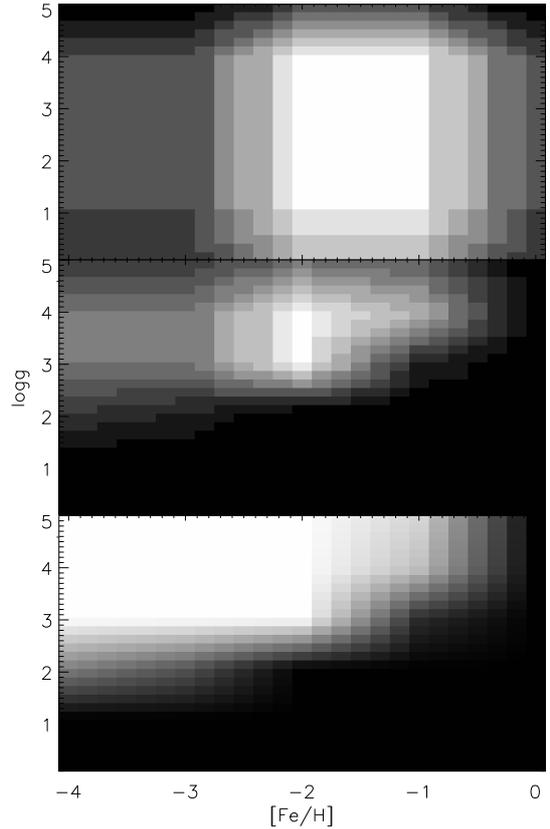}
      \caption{Analysis of the distortions in the observed metallicity
	distribution due to the BOSS color cut for the selection 
	of spectrophotometric standards. The top panel shows the distribution 
	of models under consideration (and listed in Table 1). The middle panel
	shows the distribution of models that pass the color cut in Eq. 1. 
      The bottom panel shows the ratio of the middle and top panels. White 
	indicates higher numbers on a linear scale. See
	text for details.}
         \label{distort}
   \end{figure}

    \begin{figure*}
   \centering
   \includegraphics[width=12cm,angle=90]{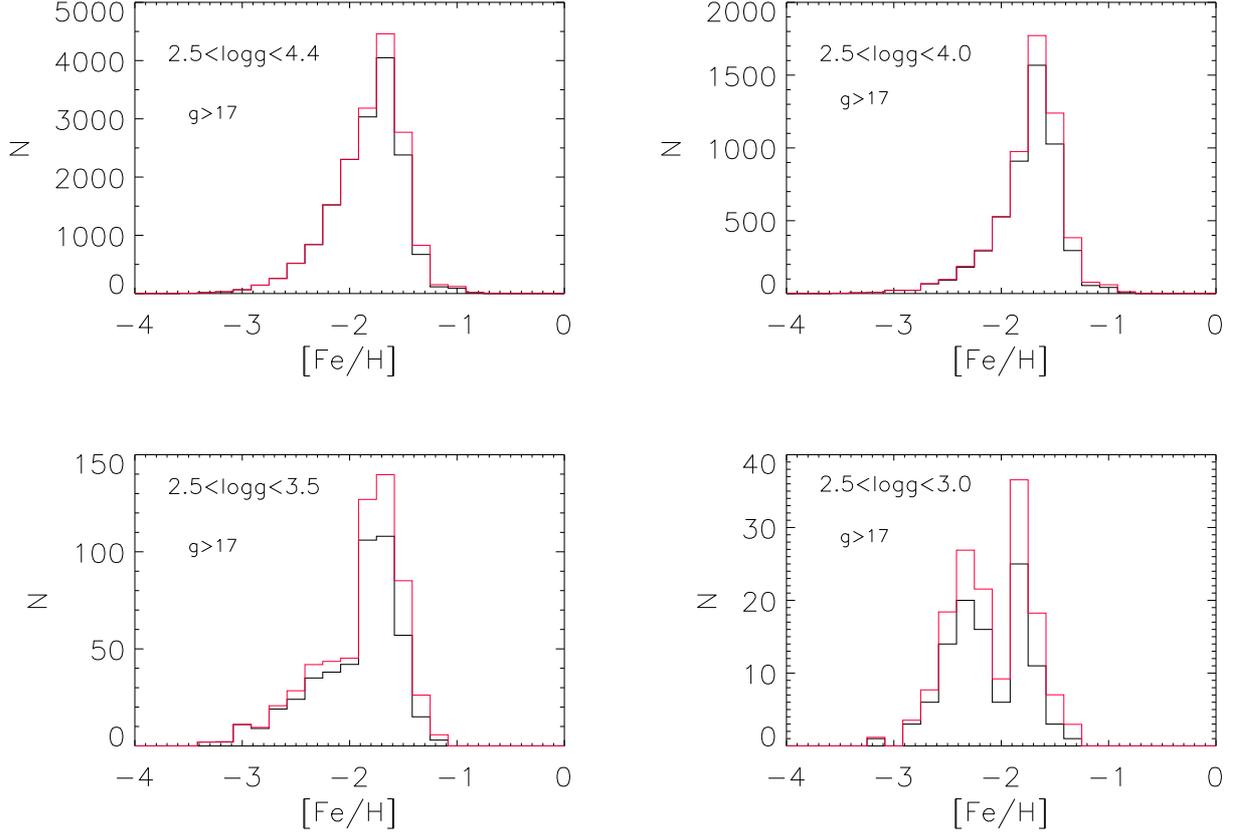}
      \caption{Metallicity distribution derived from BOSS spectra of 
      {\it spectrophotometric
      standard} stars. Black lines are as measured, and red lines after correcting 
  selections effects as explained in the text. The upper two panels show the distributions for stars with two 
different upper limits on surface gravity, $\log g$. The two lower panels illustrate the
change in metallicity distribution for stars exploring the more distant regions of the halo system.}
         \label{feh}
   \end{figure*}

The observed metallicity distribution is biased by to 
the color cuts adopted to select the BOSS spectrophotometric standard
stars, i.e., Eq. \ref{color_selection}. 
To evaluate this bias quantitatively, 
we make use of the spectral energy distributions provided with 
the very same Kurucz model atmospheres described in Section \ref{analysis}.
We calculate the SDSS $ugriz$ colors for stars with different 
atmospheric parameters, using the fiducial filter responses  (Gunn 2001, 
private communication\footnote{http://www.sdss.org/dr3/instruments/imager/\#filters}) 
for a point source at an airmass of 1.3. They correspond to measurements 
performed in 2000, but see the discussion by Doi et al. (2010) for more details 
on the time stability of the SDSS filter responses. 
The results are given in Table 1, available in electronic format.

By selecting models in Table 1 within the temperature
range $5600 \le T_{\rm eff} \le 6500$ K 
and identifying which ones would make it through the color cut in Eq. 
\ref{color_selection}, we evaluate how the color selection distorts the
underlying metallicity distribution. This is illustrated in Figure \ref{distort}.
In this figure we show 2D gray-scale maps of the distribution in 
$\log g$ and [Fe/H] for models within the range
in temperature stated above in Table 1, before (top panel) and after applying the color
selection appropriate to BOSS spectrophotometric standards (middle panel). 
These results were obtained using bins of 1.0 dex for both $\log g$ and [Fe/H],
and interpolating linearly for an improved resolution. The bottom
panel shows the ratio of the arrays depicted in the middle and top panels, 
i.e. the fraction of models that pass the color cut. 
From Figure \ref{distort} it is obvious that the color cut
used to select spectrophotometric standards is very efficient
at removing high-metallicity stars, and it becomes more
efficient at low gravities.

The resulting metallicity distribution function is depicted in 
Figure \ref{feh}. 
%All the discussed distributions are normalized for ease of comparison.
Each panel shows samples subjected to different
magnitude and gravity cuts. 
%The black lines are always as measured, and
%the red ones are corrected for the bias introduced by the color cut. 
The corrections associated with the color cut are fairly small.
The top two panels are limited to
stars with $g>17$ and either $2.5<\log g<4.4$ (top-left) or $2.5<\log g<4$ (top-right).
The more stringent gravity cut is introduced to ensure 
that all our stars are
at least 5 kpc away, minimizing contributions 
 from the thick disk or metal-weak thick disk. However, both distributions
are very similar, indicating that the  color cut is in fact very 
efficient at selecting
halo stars. In all cases we are also limiting the samples to stars 
with $T_{\rm eff}> 5600$ K and $\log \chi^2 < 0.5$. 
For the bulk of the halo, the metallicity distribution
peaks at about [Fe/H]$=-1.6$, in good agreement with previous
determinations. However, there are some striking differences with earlier results
that we discuss below. 
Table \ref{tbl-2} provides the number of stars in each
bin for the largest subsample ($g>17$ and $2.5<\log g<4.4$).

The metallicity distribution published by Allende Prieto et al. (2006),
 based on SDSS
spectra included in DR6 for  G- and F-type stars located 
at more than 8 kpc from the plane has an extended
tail toward low metallicity, 
similar to our results, but it extends to significantly 
higher metallicities. The distribution we derived is also more concentrated and asymmetric
than previous results from photometric calibrations (Ivezic et al. 2008, An et al. 2013).
Our results are perhaps more similar to those by Li et al. (2010), but 
those authors do not separate halo and thick-disk stars at high metallicity.

The two bottom panels in Fig. \ref{feh} show the metallicity 
distributions for much more restricted samples, which are limited
to stars with lower gravities, hence more distant, but still  
with $g>17$ in both cases. For $2.5<\log g<3$
(bottom right panel), we limit the sample to stars beyond 10 kpc.
In this panel we find
a clear shift of the distributions to lower metallicities, 
in line with results by Carollo et al. (2007, 2010), de Jong et al. (2010), 
and Beers et al. (2012).   The metallicity peaks at [Fe/H]$ = -1.6$ 
and [Fe/H]$= -2.2$ resemble those associated by these authors 
with the inner and outer halo of the Galaxy.

\section{Conclusions}
\label{conclusion}

We analyzed a large sample of F-type halo turnoff stellar spectra 
obtained by the SDSS-III BOSS project. We upgraded our analysis methods 
to better account for instrumental
distortions and systematic calibration errors and derived the main atmospheric
parameters, including the overall iron abundance.

Our results compare well with previously published analyses of SDSS spectra. 
The analysis of stars in the globular cluster M13 shows that the [Fe/H] estimations 
from this work have less dispersion than those provided up to now by the SSPP, 
although low-gravity metal-poor stars suffer from underestimated gravities
and metallicities.
Relative to previous work, we reach more distant regions of the 
Galactic halo with a highly homogeneous sample of F-type turnoff stars. 

Our experiments
indicate that selection effects associated with the color cuts used in SDSS
to select these stars cause no significant
distortions in the observed metallicity distributions. 
We find a halo metallicity distribution that peaks at around [Fe/H]$=-1.6$,
 in agreement with a plethora of previous studies. Nonetheless, our
distribution is more sharply concentrated, with a rapid fall-off
on the high-metallicity side, including very few stars  at [Fe/H]$>-1$
and an extended tail toward very low metallicities.

The most distant stars in our sample, at tens of kpc from the Sun,
hint at a significant change in the metallicity distribution, which
 shifts toward lower metallicities, 
consistent with expectations if the halo of the Milky Way 
comprises both an inner- and an outer-halo population. However, this has
been the subject of an intense debate in the last few years (see, e.g., 
Beers et al. 2012 vs. Schoenrich et al. 2014) which we do not pretend to 
solve in this paper, since we are focused on a small fraction of the spectra obtained by 
the SDSS. 

At this point we have only considered the overall metallicity of the
stars, as derived from all metal lines available in BOSS spectra
(380-920 mm). At the lowest metallicities and for the surface temperatures
of the F-type stars considered, the derived metallicities are dominated
by the signal in the CaII lines, whereas at higher metal abundances our metallicity
determinations are dominated by large numbers of iron lines. In 
the following papers, we will separate individual elements
and revisit the halo metallicity distribution. We will also
infer distances to the stars and examine possible 
variations in metal abundances as a function of distance from the
Galactic center in more detail.

\begin{table}
\caption{Calculated $ugriz$ colors for stars with different atmospheric parameters.
 Full table available in electronic form \label{tbl-1}}
\centering
\begin{tabular}{ccccccc}
\hline\hline
[Fe/H] & $T_{\rm eff}$ & $\log g$ & $(u-g)_0$ & $(g-r)_0$ & $(r-i)_0$ & $(i-z)_0$ \\
\hline
\hline
0.5  &  3500  &  0.00  &  3.873  &  1.472  &  1.308  &  0.717\\
0.5  &  3500  &  0.50  &  3.519  &  1.374  &  1.402  &  0.752\\
0.5  &  3500  &  1.00  &  3.176  &  1.288  &  1.477  &  0.782\\
0.5  &  3500  &  1.50  &  2.845  &  1.215  &  1.533  &  0.805\\
\dots & \dots & \dots & \dots & \dots & \dots & \dots \\
\end{tabular}                                                
\end{table}

\begin{table}
\caption{Metallicity distribution derived from BOSS spectra of spectrophotometric
standards with $T_{\rm eff} > 5600$ K, $2.5 \le \log g \le 4.4$, and $g>17$. 
All metallicity bins are 0.1 dex wide. \label{tbl-2}}
\centering
\begin{tabular}{ccc}
\hline\hline
[Fe/H] (bin center) & N & N/Ntotal  \\
\hline
\hline   
-4.000&     1&  0.0001 \\
-3.833&     1&  0.0001 \\
-3.667&     0&  0.0000 \\
-3.500&     5&  0.0003 \\
-3.333&    16&  0.0009 \\
-3.167&    30&  0.0017 \\
-3.000&    65&  0.0038 \\
-2.833&   144&  0.0084 \\
-2.667&   260&  0.0151 \\
-2.500&   521&  0.0302 \\
-2.333&   845&  0.0490 \\
-2.167&  1525&  0.0884 \\
-2.000&  2305&  0.1336 \\
-1.833&  3183&  0.1845 \\
-1.667&  4459&  0.2584 \\
-1.500&  2768&  0.1605 \\
-1.333&   828&  0.0480 \\
-1.167&   152&  0.0088 \\
-1.000&   121&  0.0070 \\
-0.833&    21&  0.0013 \\
-0.667&     0&  0.0000 \\
-0.500&     0&  0.0000 \\
-0.333&     0&  0.0000 \\
-0.167&     0&  0.0000 \\
 0.000&     0&  0.0000 \\
\hline
\end{tabular} 
\end{table}

\begin{acknowledgements}

We are grateful to Ivan Hubeny and Lars Koesterke for help with calculations
of opacities and model spectra, to David Schlegel for stimulating 
discussions and advice, and Steph Snedden and Howard Brewington for their 
observations.
TCB acknowledges partial support for this work by grant PHY 08-22648: 
Physics Frontiers Center/Joint Institute for Nuclear Astrophysics (JINA), 
awarded by the U.S. National Science Foundation. YSL is
a Tombaugh Fellow at New Mexico State University.

Funding for SDSS-III has been provided by the Alfred P. Sloan Foundation, 
the Participating Institutions, the National Science Foundation, and the U.S. 
Department of Energy Office of Science. The SDSS-III web site is http://www.sdss3.org/.

SDSS-III is managed by the Astrophysical Research Consortium for the 
Participating Institutions of the SDSS-III Collaboration including the 
University of Arizona, the Brazilian Participation Group, Brookhaven National Laboratory, 
University of Cambridge, Carnegie Mellon University, University of Florida, 
the French Participation Group, the German Participation Group, Harvard University, 
the Instituto de Astrofisica de Canarias, the Michigan State/Notre Dame/JINA 
Participation Group, Johns Hopkins University, Lawrence Berkeley National Laboratory, 
Max Planck Institute for Astrophysics, Max Planck Institute for Extraterrestrial Physics, 
New Mexico State University, New York University, Ohio State University, 
Pennsylvania State University, University of Portsmouth, Princeton University, 
the Spanish Participation Group, University of Tokyo, University of Utah, 
Vanderbilt University, University of Virginia, University of Washington, 
and Yale University. 

\end{acknowledgements}

\end{document}